\documentclass[twocolumn]{aastex631}
\usepackage{amsmath}
\usepackage{comment}

\begin{document}

\title{Pair creation in hot electrosphere of compact astrophysical objects}

\author{Mikalai Prakapenia}
\affiliation{ICRANet-Minsk, Institute of Physics, National Academy of Sciences of Belarus, 220072 Nezalezhnasci Av. 68-2, Minsk, Belarus}
\affiliation{Department of Theoretical Physics and Astrophysics, Belarusian State University, Nezalezhnasci Av. 4, 220030 Minsk, Belarus}

\author{Gregory Vereshchagin}
\affiliation{ICRANet-Minsk, Institute of Physics, National Academy of Sciences of Belarus, 220072 Nezalezhnasci Av. 68-2, Minsk, Belarus}
\affiliation{ICRANet, 65122 Piazza della Repubblica, 10, Pescara, Italy}
\affiliation{ICRA, Dipartimento di Fisica, Sapienza Universit\`a di Roma, Piazzale Aldo Moro 5, I-00185 Rome, Italy}
\affiliation{INAF -- Istituto di Astrofisica e Planetologia Spaziali, 00133 Via del Fosso del Cavaliere, 100, Rome, Italy}

\begin{abstract}
The mechanism of pair creation in electrosphere of compact astrophysical objects such as quark stars or neutron stars is revisited, paying attention to evaporation of electrons and acceleration of electrons and positrons, previously not addressed in the literature. We perform a series of numerical simulations using the Vlasov-Maxwell equations. The rate of pair creation strongly depends on electric field strength in the electrosphere. Despite Pauli blocking is explicitly taken into account, we find no exponential suppression of the pair creation rate at low temperatures. The luminosity in pairs increases with temperature and it may reach up to $L_\pm\sim 10^{52}$ erg/s, much larger than previously assumed.
\end{abstract}



\section{Introduction}

Quantum electrodynamics predicts vacuum breakdown in strong electric field, leading to prolific creation of electron-positron pairs. An extreme value of electric field strength required for this process $E\sim E_c=m_{e}^2c^3/\hbar e \simeq 1.3\times 10^{16}$ V/cm, where $m_{e}$ is the electron mass, $e$ is its charge, $c$ is the speed of light and $\hbar$ is reduced Planck constant, is not yet reachable in laboratory conditions. However, one may look for this process in some extreme astrophysical environments, for reviews see \cite{2010PhR...487....1R,2022Univ....8..473V}. Indeed, strong electric fields may exist on bare surfaces of hypothetical quark stars \citep{1986ApJ...310..261A,1995PhRvD..51.1440K,1998PhRvL..80..230U,2005ApJ...620..915U,2006ApJ...643..318H,2010PhRvD..82j3010P,2023arXiv231112511I} or even neutron stars \citep{2011PhLB..701..667R,2011PhRvC..83d5805R,2012NuPhA.883....1B,2014PhRvC..89c5804R}. The region with overcritical $E>E_c$ electric field in these objects is called \emph{electrosphere}. Similar electrospheres are predicted for such hypothetical objects as superheavy nuclei \citep{1976JETPL..24..163M} and quark nuggets \citep{2010PhRvD..82h3510F}. The magnitude of electric field in electrosphere depends on the sharpness of the boundary of positively charged component \citep{2010JPhG...37g5201M}. \citet{1998PhRvL..80..230U} in his seminal paper proposed that hot quark stars may be a source of pair winds, potentially observable at cosmological distances. Based on this work detailed study of particle interactions was performed by \citet{2004ApJ...609..363A,2005ApJ...632..567A} predicting observed properties of hot quark stars.

In this paper we revisit Usov's mechanism of pair creation in electrosphere of compact objects. First, we show that the reasoning under Usov's results contain some flaws. Then, we provide new arguments how pair creation can operate, and derive the rate of pair creation together with pair luminosity for electrosphere of a compact astrophysical object. The self-consistent simulations for electron-positron pair creation and electric field evolution in electrosphere are performed. Our results indicate that hot electrosphere indeed is a source of strong pair wind. The paper is organized as follows. In section \ref{secmechanism} we describe the original Usov's idea. In section \ref{secrate} we derive the rate of pair creation in electrosphere. In section \ref{dynamics} we present main equations and corresponding boundary conditions used for simulations. Our numerical results are presented in section \ref{numres}. Discussion and conclusions follow in sections \ref{diss} and \ref{concl}, respectively.

\section{Usov's mechanism}
\label{secmechanism}

\citet{1998PhRvL..80..230U} was the first who pointed out the possibility of pair creation in electrosphere of compact astrophysical objects. 
Since the work by \citet{1986ApJ...310..261A} it is argued that near the surface of bare quark star a supercritical electric field may exist. This is because the sharpness of the quark star surface is determined by the strong interactions, while degenerate electrons are bound to quarks by electromagnetic interactions. Thus electron spatial distribution extends to larger, still microscopic, distances. The resulting charged layer generates strong and overcritical electric field. While overcritical electric field in vacuum should produce electron-positron pairs, this does not happen at small temperatures. The reason is that all electronic states in this configuration are fully occupied and the Schwinger process is forbidden. In order to start pair creation process empty electronic states have to be present. This is possible when the quark star is just formed, being very hot with its surface temperature $k_B T_S\sim \varepsilon_{F}\sim 20$ MeV \cite{1998PhRvL..80..230U}, where $\varepsilon_F$ is Fermi energy inside the star, $T_S$ is the temperature of the star, $k_B$ is the Boltzmann constant.

The mechanism proposed by \citet{1998PhRvL..80..230U} is the following. Given that the rate of pair creation in overcritical electric field is extremely fast, all empty states below the pair creation threshold are instantly occupied by creating electrons and positrons. Then the slower process of thermalization of electrons determines the appearance of new empty states for electrons. Positrons are ejected by the electric field. Their outflow leads to outflow of electrons. The number of created electron-positron pairs $N_\pm$ per unit volume $\Delta V$ and unit time is estimated as \citep{1998PhRvL..80..230U}:
\begin{equation}
\label{usovrate}
\dot n_\pm = \frac{\dot N_\pm}{S_R \Delta r_E} \simeq \Delta n_e /\tau_{ee},
\end{equation}
where $S_R$ is the surface area of electrosphere, $\Delta r_E$ is its thickness, $\tau_{ee}$ is the thermalization timescale of electron-electron collisions, and a dot denotes time derivative.
The density of electronic empty states below the pair creation threshold $\Delta n_e$ in the strong degeneracy approximation $k_B T_S \ll \varepsilon_F$ is
\begin{equation}
\Delta n_e \simeq \frac{3k_B T_S}{\varepsilon_F}n_e \exp{\left(-\frac{2m_e c^2}{k_B T_S}\right)},
\label{densityofstates}
\end{equation}
the thermalization timescale $\tau_{ee}$ is given by
\begin{equation}
\tau_{ee}^{-1}\simeq \frac{3\alpha}{2\pi^{3/2}}\frac{(k_B T_S)^2}{\hbar \varepsilon_F}J(\zeta),
\end{equation}
where $\zeta=2\alpha^{1/2}\pi^{-1/2}\frac{\varepsilon_F}{k_B T_S}$, $\alpha$ is the fine structure constant and
\begin{equation}
J(\zeta)=\left\{
\begin{array}[c]{cc}
51(1-19.5 \zeta^{-1}+296\zeta^{-2}), & \zeta>20,\\
0.23\zeta^{1.8} , & 1<\zeta<20, \\
(1/3)\zeta^3\ln{2/\zeta}, & \zeta<1.
\end{array}
\right.
\end{equation}

The thickness of the electrosphere $\Delta r_E$ is given by \cite{2005ApJ...620..915U}
\begin{equation}
\Delta r_E = \left(\frac{3\pi}{2\alpha}\right)^{1/2}\frac{m_e c^2}{e\varphi_0}\frac{\hbar}{m_e c},
\end{equation}
where $\varphi_0$ is electrostatic potential on the surface of the compact object. For the case of fully degenerate electrons we have $e\varphi_0 = \frac{3}{4}\varepsilon_F$. Thus, for $\varepsilon_F=20$ MeV one finds $\Delta r_E \simeq 3\times10^{-11}$ cm. 

It is found that the product $\Delta r_E \Delta n_e$ is independent on the temperature. In addition, since the number of created pairs is considerably smaller than the number of electrons in electrosphere, it is assumed that the structure of electrosphere is not modified by the process of pair creation. Therefore, the key assumption adopted by \citet{1998PhRvL..80..230U,2005ApJ...620..915U} is that electric field acts as a catalyst for the Schwinger process and it does not affect particles dynamics.

According to the Usov mechanism, the process of pair creation continues as long as the temperature remains high enough, $k_B T_S > 2 m_e c^2$, see eq. (\ref{densityofstates}). At lower temperatures pair creation is exponentially suppressed.

\section{Pair creation rate in electrosphere}
\label{secrate}

Assuming pairs are not produced due to occupied electron states, one can find a static solution for the Vlasov-Maxwell equations. Strictly speaking this is true only for fully degenerate electrons. One may assume that quasi-static equilibrium is also possible\footnote{We use the term quasi-static, because at nonzero temperature thermal evaporation of electrons modifies the structure of electrosphere, see below.} for nonzero temperatures \citep{1995PhRvD..51.1440K}. Electrons in equilibrium obey the Fermi-Dirac statistics with their $f_e$ distribution function
\begin{equation}\label{FDdistr}
f_e= \frac{1}{1+\exp{\left[\left(\sqrt{{p}^{2} + m_e^2}-{\mu}\right)/ T_S\right]}},
\end{equation}
where $\mu$ is their chemical potential\footnote{In this section and below we use the units with $\hbar=c=k_B=1$.}. The chemical equilibrium condition for electrons is
\begin{equation}
\label{chemeq}
\mu=e\varphi,
\end{equation}
implying that electrons are bound by the electrosphere and also that their distribution function does not depend on time $\partial f_e/ \partial t =0$.
The number density of electrons for $T_s\ll \mu$ is \citep{1995PhRvD..51.1440K,2005ApJ...620..915U}
\begin{equation}
    n_e=\frac{\mu^3}{3\pi^2}+\frac{\mu T_S^2}{3}.
\end{equation}

\begin{figure}[h]
\includegraphics[width=\columnwidth]{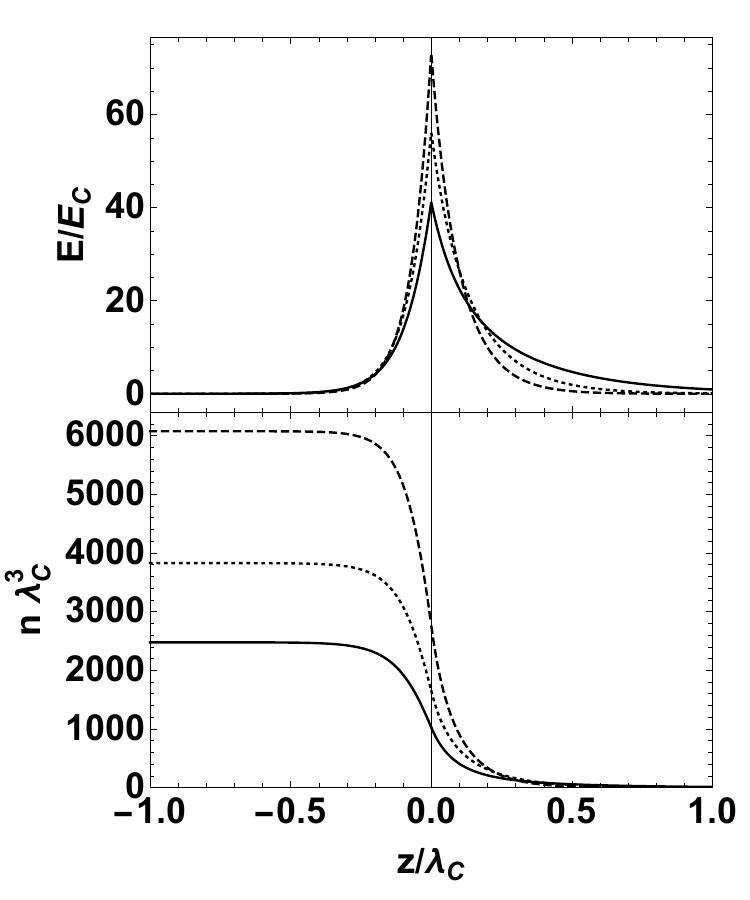}
\caption{Electric field (top) and electron density (bottom) spatial distribution for selected temperatures $T_S$: $T_S=3m_e$ (solid), $T_S=6m_e$ (dotted),$T_S=9m_e$ (dashed). Here $e \varphi_q=20$ MeV and $\lambda_C\simeq 2.4\times 10^{-10}$ cm is electron Compton wavelength.}
\label{figpoisson}
\end{figure}

In ultrarelativistic approximation for electrons the Poisson equation gives \citep{1986ApJ...310..261A,1995PhRvD..51.1440K}
\begin{equation}\label{poisson}
    \frac{d^2\varphi}{dz^2}=-\frac{4\alpha}{3\pi}\left[e^2\varphi^3+\pi^2 T_s^2\varphi-n_q\right],
\end{equation}
where $n_q(z<0)=(\alpha/3\pi^2) (e^2\varphi_q^3+\pi^2 T_S \varphi_q)$ is the density of a positively charged core, $z$ is spatial coordinate normal to electrosphere. The electric field $E(z)$ is defined from the electrostatic potential $\varphi(z)$ as $E(z)=-d\varphi/dz$. The solution of the electrostatic problem is shown in Fig. \ref{figpoisson} for selected values of temperature. At the surface $z=0$ the values of electric field $E_0$ and electrostatic potential $\varphi_0$ can be expressed as \citep{1995PhRvD..51.1440K,2005ApJ...620..915U}:
\begin{gather}\label{EmaxofT}
E_0 = \sqrt{\frac{2\alpha}{3\pi}}\sqrt{e^4\varphi^4_0+2\pi^2T^2e^2\varphi_0^2},\\
e\varphi_0= \frac{\varepsilon_F(2\pi^2T^2+3\varepsilon_F^2)}{4(\pi^2T^2+\varepsilon_F^2)}.
\end{gather}

It is important to note that this solution is not fully self-consistent. It ignores thermal evaporation of electrons, and the consequent electrosphere \emph{inflation}. The condition of chemical equilibrium (\ref{chemeq}) is valid at zero temperature. At non-zero temperatures some electrons have large enough energies to overcome the Coulomb barrier and leave the electrosphere. Therefore, spatial distribution of electrons as well as electric field in hot electrosphere can extend on much larger distances from the surface, than in the cold case. This effect is not accounted for by eq. (\ref{poisson}) and is not shown in Fig. \ref{figpoisson}. However, our numerical results clearly demonstrate this effect, see below.

The basic assumption made by \citet{1998PhRvL..80..230U} is that the role of the electrosphere is just creation of electron-positron pairs out of electric field. Their subsequent evolution is determined by collisions, leading to thermalization. In this approach particle acceleration by the electric field is neglected. While thermalization is indeed relevant process for reaching equilibrium in relativistic plasma \citep{2007PhRvL..99l5003A,2009PhRvD..79d3008A,2010PhRvE..81d6401A}, there are other kinetic processes that may change electron distribution function. Below we demonstrate that within electrosphere pairs are practically collisionless.

The Coulomb logarithm $\Lambda$ in relativistic plasma is of the order of unity \citep{2017rkt..book.....V}. Hence the transport cross section for Coulomb collisions of electrons is $\sigma_\text{coul}=\sigma_T \left(m_e/T_S \right)^2 \Lambda \sim \sigma_T$, where $\sigma_T$ is the Thomson cross-section. Thus, the Thomson cross-section can be used as an estimation for the Coulomb scattering cross-section. The mean free path $l$ is 
\begin{equation}
\label{mfp}
l\simeq(\sigma_T n_e)^{-1}=\frac{3}{8\pi\alpha^2}\left(\frac{n_e}{\lambda_C^{-3}}\right)^{-1}.
\end{equation} 
Assuming that electron density in the electrosphere is $n_e\sim10^{35}~\text{cm}^{-3}$ \citep{1998PhRvL..80..230U}, see Fig. \ref{figpoisson}, we obtain $l\sim 10^{-11}$ cm. From these estimations we see that the thickness of cold electrosphere is equal to the mean free path of particles. As the density of electrons decreases exponentially with radius, it  makes the entire region above electrosphere almost collisionless.

We can estimate the rate of pair creation using the Schwinger rate per unit volume and per unit time in a constant in time and homogeneous electric field \cite{1951PhRv...82..664S,1970SJMP...11..596N}
\begin{equation}\label{schwingerW}
\dot n_\pm = \frac{m_e^4}{4\pi^3}\left(\frac{E}{E_c}\right)^2 \exp{\left( -\pi \frac{E_c}{E} \right)}.
\end{equation}

To describe pair creation for non-vacuum initial state it is necessary to take into account Pauli blocking effect \citep{2023PhRvD.108a3002P,2023PhRvE.107c5204B}. We use the differential pair creation rate given by \citet{1987PhRvD..36..114G} with additional Pauli blocking factor $(1-f_e)$. The rate is then given by an integral over particle momentum $d^3p_e$ in the following form
\begin{gather}
\label{Paulirate}
\dot n_\pm = - \frac{2\pi m_e^4}{4\pi^3} \frac{E}{E_c}
\int\limits_0^\infty d\tilde p \tilde p \left(1-f_e\right) \\ \notag
\times \ln\left[  1-\exp{\left(-\frac{\pi(\tilde p^2+1)}{E/E_c} \right)}  \right]. 
\end{gather}
Without the Pauli blocking factor $(1-f_e)$ this expression reduces to the Schwinger rate \eqref{schwingerW}. Therefore, the rate depends on three parameters: temperature $T_S$, chemical potential $\mu$ and electric field $E$. The electric field and chemical potential can be obtained from the electrostatic configuration as a functions of the temperature.
The only remaining free parameter is the temperature. 

\begin{figure}[h]
\includegraphics[width=\columnwidth]{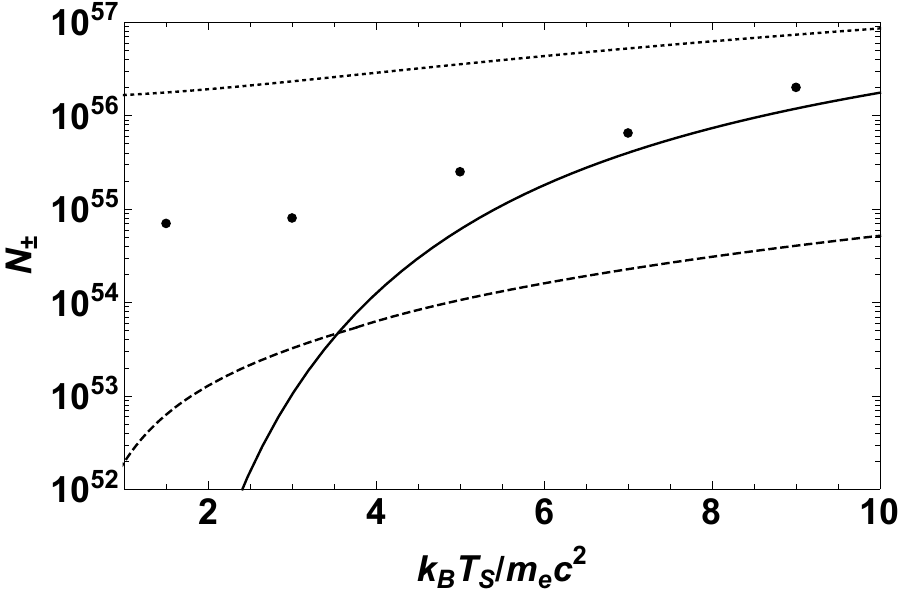}
\caption{Pair creation rate in the electrosphere $\dot N_\pm$ according to eq. (\ref{usovrate}) for dashed curve, eq. \eqref{maxrate} for dotted curve and eq. (\ref{Paulirate}) for solid curve. Dots represent numerical results from the Tab. \ref{tab1}.}
\label{fig0}
\end{figure}

The total rates of pair creation can be computed using the typical parameters for strange stars: $R=10^6$ cm, $\varepsilon_F=20$ MeV and $\Delta r_E=10^{-10}$ cm \citep{1986ApJ...310..261A,1998PhRvL..80..230U}. We show this quantity in Fig. \ref{fig0}: dashed line represents Usov's rate obtained using eq. (\ref{usovrate}), while solid line represents the rate computed with eq. (\ref{Paulirate}) and dotted line corresponds to eq. \eqref{maxrate}, where the surface quantities \eqref{EmaxofT} were substituted.

At high temperatures $T_S>\varepsilon_F$ Pauli blocking becomes negligible and since $E\gg E_c$ one finds \citep{1998PhRvL..80..230U}
\begin{gather}
\label{maxrate}
\dot N_\pm \simeq 4\pi R^2 \Delta r_E \frac{m_e^4}{4\pi^3}\left(\frac{E}{E_c}\right)^2 \\ \notag 
\simeq 4\times 10^{56}\; {\rm s}^{-1} \left(\frac{E}{5\times 10^{17} {\rm V/cm}}\right)^2.
\end{gather}

We find that the luminosity in pairs is determined not by their temperature, as Usov assumed, but by the Lorentz factor of the outflow. Positrons are accelerated by the electrosphere and ejected with high Lorentz factor, which can be estimated from the equation of motion \begin{equation}
\gamma\simeq \frac{E}{E_c}.
\label{gamma}
\end{equation}
Electrons, while decelerated in the electrosphere, are dragged together with positrons. In fact, the flux of positrons reduces the Coulomb barrier thereby allowing more electrons to escape. This guarantees charge neutrality of the total outflow. From (\ref{maxrate}) and (\ref{gamma}) we find that the luminosity in pairs can be as large as
\begin{gather}
\label{maxlum}
L_\pm \simeq 1.3\times 10^{52}\; {\rm erg/s} \left(\frac{E}{5\times 10^{17} {\rm V/cm}}\right)^3.
\end{gather}
In this estimate the reference value of electric field $E=30E_c$, see Fig. \ref{figpoisson} in the electrosphere is used. Given strong dependence on eletric field even larger values are expected.
This result shows that even very large isotropic luminosities of gamma-ray bursts (GRBs) of $10^{53}$ erg/s and higher are possible for electrosphere of compact objects, provided that $E\sim 80 E_c$ or higher at their surface.

The source of energy for pair outflow is thermal energy of the compact object which can be as high as $10^{53}$ erg \citep{1991ApJ...375..209H} for the temperatures of $10^{11}$ K.

Note, that in the derivation of eq. (\ref{Paulirate}) the role of Pauli blocking is overestimated. Pair acceleration by the electric field is so strong that operates as an alternative mechanism to thermalization by creating empty states in the phase space and thus allowing pair creation to operate. In order for this mechanism to work the rate of pair creation obtained from (\ref{schwingerW}) should be smaller than the rate of acceleration, obtained from (\ref{gamma}), which leads to the following constraint \citep{2011PhLB..698...75B}
\begin{equation}
\frac{1}{4\pi^{3}}\frac{E}{E_c}\exp\left(  -\pi\frac{E_c}{E}\right)
\leq 1,\label{blockingrate}%
\end{equation}
which gives $E\lesssim 127E_{c}$. Therefore, for typical values of electric field in electrosphere this effect strongly enhances pair creation.

\section{Dynamical equations and initial conditions at the surface}
\label{dynamics}

As we consider particle dynamics in orthogonal direction to the surface of the compact object we introduce cylindrical coordinates in momentum space $\mathbf{p}%
=\{p_{\perp },p_\phi ,p_{||}\}$ with $p_{||}$-axis parallel to electric field $E$. Particle energy is then $p^{0}=[p_{\perp }^{2}+p_{||}^{2}+m_{e}^{2}]^{1/2}$.

Particle evolution is described by one-particle electron/positron distribution function $f_\pm(t,z,p_{\perp },p_{||})$, which is normalized on
particle density $n_\pm=\int \frac{d^{3}p}{(2\pi )^{3}}f_\pm$.

We introduce dimensionless quantities $\tilde{t}=tm$, $\tilde{z}=zm$, $\tilde{p} =p/m$, $\tilde{E}=Ee/m_{e}^{2}$ to write the basic dynamic equations in dimensionless form. The set of Maxwell-Vlasov equations in our case reduces to the Vlasov-Amp\'ere system, which has to be supplemented with the Gauss law, as an initial condition. For this purpose we use the Poisson equation \eqref{poisson}. We have
\begin{gather}
\label{vlasovampere}
\frac{\partial f_\pm}{\partial \tilde t} +\frac{\tilde p_{||}}{\tilde p^0}\frac{\partial f_\pm}{\partial \tilde z}\mp \tilde E\frac{\partial f_\pm}{\partial \tilde p_{||}}= \\ \notag
-(1-f_{+}-f_{-})|\tilde E|\text{ln}\left[1-\exp\left(  -\frac{\pi(1+\tilde p_\perp^2)}{|E|}\right)\right] \delta(\tilde p),   \\ \notag
\frac{\partial E}{\partial t} = 2 \alpha\int d^3 \tilde p \frac{\tilde p}{\tilde p^0}(f_- - f_+)+ \\ \notag
4 \alpha\frac{|\tilde E|}{\tilde E} (1-f_{+}-f_{-})\text{ln}\left[  1-\exp\left(-\frac{\pi(1+\tilde p_\perp^2)}{|\tilde E|}\right)  \right]\delta(\tilde p). 
\end{gather}

We recall that the solution of the Poisson equation \eqref{poisson} leads to the static solution of the Vlasov-Amp\'ere system \eqref{vlasovampere} only in the case $T_S=0$, that is for fully degenerate electrons. Only in this case the chemical potential $\mu=e \varphi$ equals Fermi energy, which means that there are no electrons with energies exceeding the Coulomb barrier. However, when the surface temperature is non-zero $T_S>0$ there is a small part of electrons with energies larger than the Coulomb barrier. These electrons move outward the surface increasing the electric field outside the surface of the compact object, leading to electrosphere inflation.

\begin{figure}[ht!]
\includegraphics[width=\columnwidth]{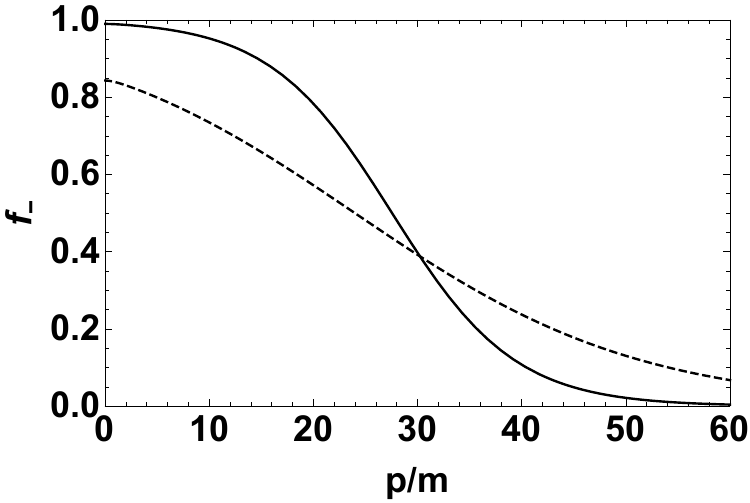}
\caption{Initial distribution of electrons in the momentum space at the surface $z=0$ of the compact object for different surface temperatures: $T_S=3m_{e}$ (solid) and $T_S=7m_{e}$ (dashed).}\label{figinitf}
\end{figure}

The initial distribution function of electrons is assumed to be the Fermi-Dirac one \eqref{FDdistr}, where the chemical potential $\mu=e\varphi$ and the temperature $T_S$ are obtained from the Poisson equation \eqref{poisson}. The initial distribution function of electrons at the surface $z=0$ is shown in Fig. \ref{figinitf} for two different values of temperature. The smaller is the temperature, the larger is degeneracy of electrons at small momenta, where pair creation operates, and one can expect much less pairs in this case, see Fig. \ref{fig0}.



\section{Numerical results}
\label{numres}

\begin{figure}[h]
\includegraphics[width=\columnwidth]{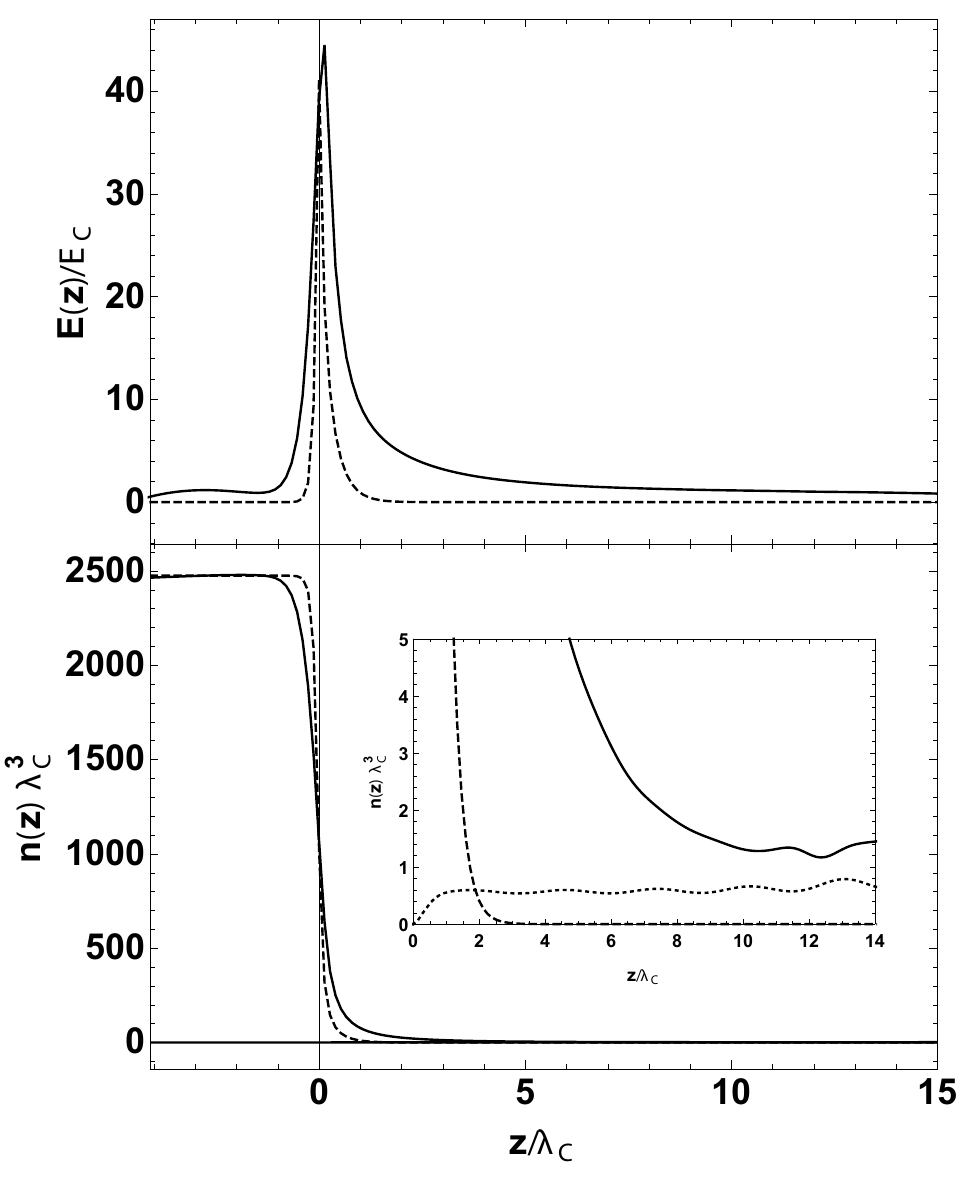}
\caption{Electric field (top) and electron number density (bottom) as function of distance from the surface of the compact object. Dashed curves represent electrostatic solution. Solid curves represents inflated electrosphere. Inset shows electron (solid) and positron (dotted) spatial distribution. Here $T_S=3m_e$.}
\label{fig1}
\end{figure}

\begin{figure}[h]
\includegraphics[width=\columnwidth]{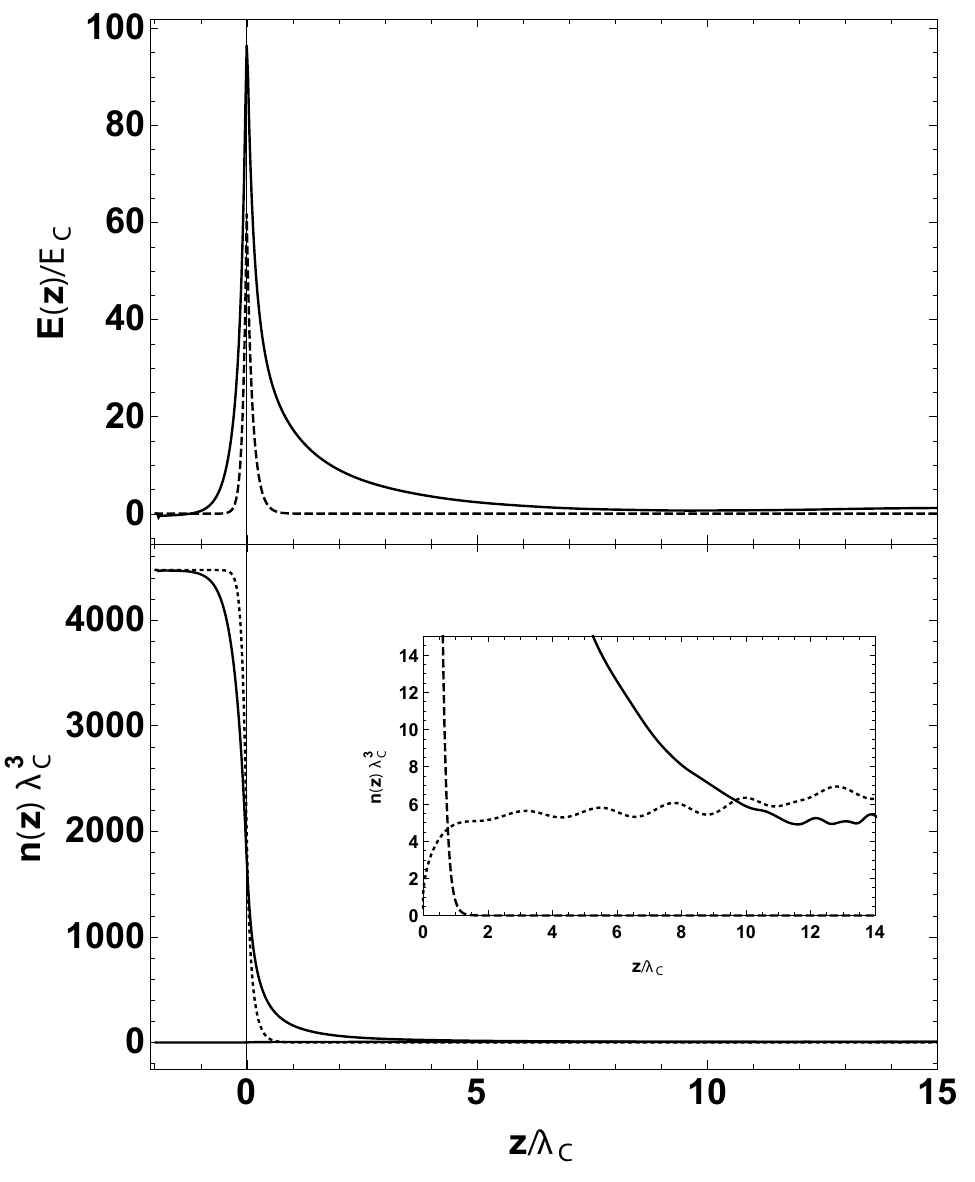}
\caption{The same as in Fig. \ref{fig1} for $T_S=7m_e$.}
\label{fig2}
\end{figure}

We integrate numerically the system of equations (\ref{vlasovampere}) and present the results for $T_S=3m_e$ in Fig. \ref{fig1}, and for $T_S=7m_e$ in Fig. \ref{fig2}. 
Electric field and electron number density as function of distance from the surface of the compact object are shown. Electrostatic solutions of eq. (\ref{poisson}), used as initial conditions for simulations, are represented by dashed curves. Solutions of dynamical equations (\ref{vlasovampere}) are shown by solid curves. Electrosphere inflation due to evaporation of electrons, whose energy exceeds the Coulomb barrier leads to extension of the region of overcritical electric field from $z\sim \lambda_C$ up to $z\sim 10\lambda_C$. Electron-positron pairs are mostly produced near the surface at $z=0$, where electric field is the largest. The combined flux of electrons and positrons is clearly visible in the insets in Figs. \ref{fig1} and \ref{fig2} at $z>10\lambda_C$.

\begin{figure}[ht!]
\includegraphics[width=\columnwidth]{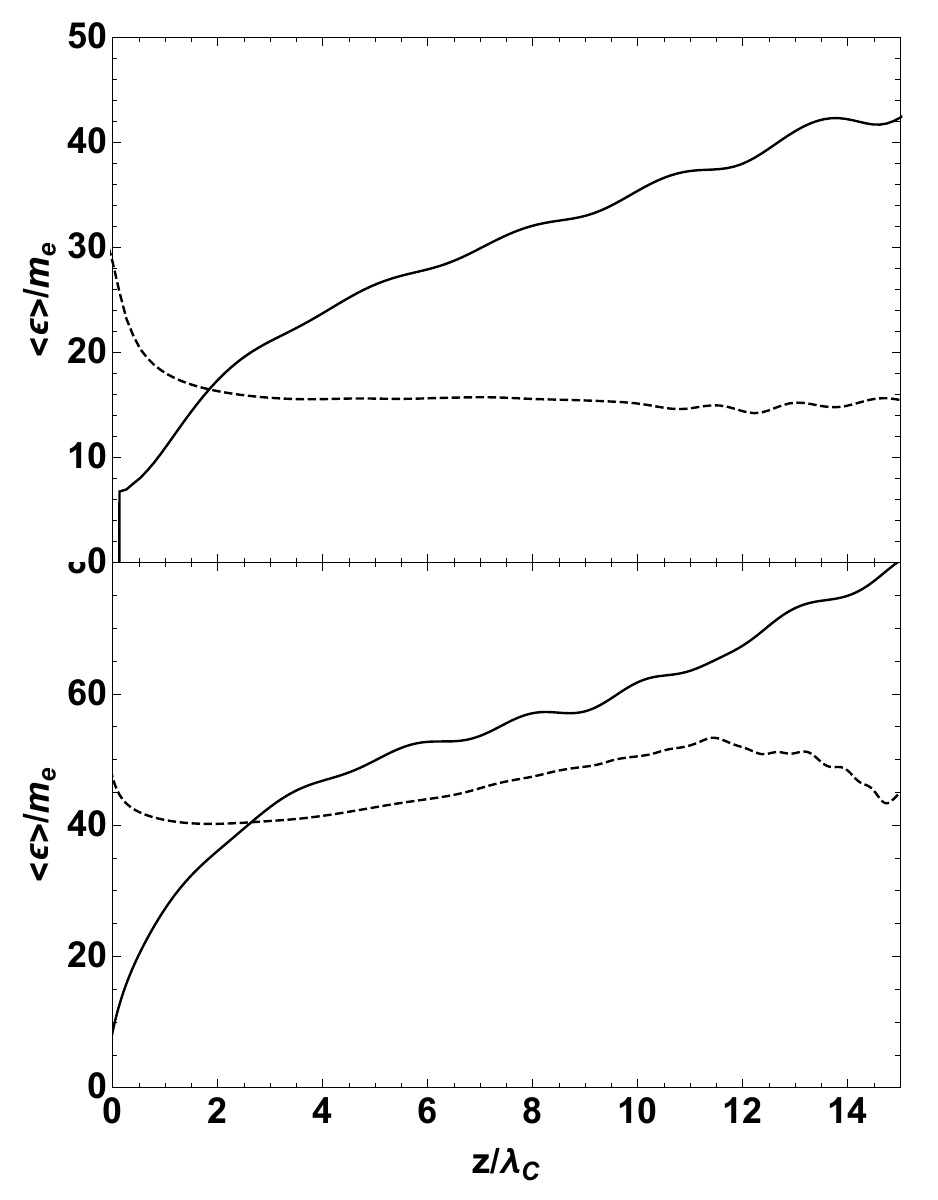}
\caption{Average energy of electrons (dashed) and positrons (solid) for $T_S=3m_e$ (top) and $T_S=7m_e$ (bottom) as a function of distance from the surface.}
\label{fig3}
\end{figure}

The average energy of electrons and positrons is computed as
\begin{equation}
\langle \tilde\varepsilon_\pm \rangle = \frac{1}{\tilde n_{\pm}}\int 2\frac{d^3\tilde p}{(2\pi)^3} \tilde p^0 f_\pm.
\end{equation}
In Fig. \ref{fig3} we present this quantity as a function of distance from the surface. It is clear that the Coulomb barrier located as small $z$ decelerates electrons and accelerates positrons. At larger distances positrons are additionally accelerated by the inflated electrosphere. The Lorentz factor $\gamma$ of the outflow here is approximately equal to the average energy. Specifically we find $\gamma\simeq 40$ for $T_S=3m_e$ and $\gamma\simeq 50$ for $T_S=7m_e$, in agreement with the estimate (\ref{gamma}).

Finally, we compute the luminosity of the outflow at a distance $\tilde z$ from the boundary as follows
\begin{equation}
\tilde L = 4\pi (\tilde R+\tilde z)^2 \int 2\frac{d^3\tilde p}{(2\pi)^3} \tilde p_{||} (f_+ + f_-)
\end{equation}
The peak luminosity for $R=10$ km and $T_S=3m_e$ is $L\simeq 2\times 10^{50}$ erg/s. For $T_S=7m_e$ we obtain $L\simeq 4\times 10^{51}$ erg/s. The values $\dot N\simeq L/\langle \tilde\varepsilon_\pm \rangle$ are $6\times 10^{54}\,\rm{s}^{-1}$ and $10^{56}\,\rm{s}^{-1}$, respectively, see Fig. \ref{fig0}.

Note, that the average density of pairs for $T_S=3m_e$ is about 1 pair in Compton volume, namely $n_\pm\simeq 7\times 10^{28}$ cm$^{-3}$. For $T_S=7m_e$ we find $n_\pm\simeq 4\times 10^{29}$ cm$^{-3}$. These densities are much smaller than the density of electrons at the boundary $z=0$. The mean free path (\ref{mfp}) is $2\times 10^3\lambda_C$ and $1.2\times 10^4\lambda_C$ respectively, much larger than the Compton length. On such distances from the boundary the outflow is collisionless and does not thermalize. We also performed simulations for $T_S=1.5m_e$, $T_S=5m_e$ and $T_S=9m_e$. All results are collected in the Tab. \ref{tab1}.
\begin{table}[tbp]
\centering%
\begin{tabular}[t]{|c|c|c|c|c|c|}
\hline
$T_S$                & $1.5m_e$  &  $3m_e$    &  $5m_e$  & $7m_e$ &  $9m_e$  \\ \hline\hline
$\dot N_\pm,~s^{-1}$ & $7\times10^{54}$ &  $8\times10^{54}$ &   $3\times10^{55}$    &  $6\times10^{55}$ &  $2\times10^{56}$   \\ \hline
$L_\pm,~\text{erg/s}$& $1\times10^{50}$ &  $2\times10^{50}$ &  $1\times10^{51}$   & $4\times10^{51}$ &  $1\times10^{52}$ \\ \hline
$\langle \tilde\varepsilon_- \rangle$           &  $8$ &  $16$ &   $27$    &  $46$ &  $68$   \\ \hline
$\langle \tilde\varepsilon_+ \rangle$           & $12$ &  $30$ &  $45$      & $54$  &  $65$ \\ \hline
\end{tabular}%
\caption{Average pair creation rate $\dot N_\pm$, average luminosity $L_\pm$, average energy per particle of electrons $\langle \tilde\varepsilon_- \rangle$ and positrons $\langle \tilde\varepsilon_+ \rangle$ obtained from numerical simulations at $t=17 \lambda_C/c$ for selected temperatures $T_S$.}
\label{tab1}
\end{table}
The numerical rate of pair creation is also represented in Fig. \ref{fig0}. 
It is clear that the values of pair creation rate and luminosity exceed the estimates based on Usov's formula (\ref{usovrate}) and the estimates, which assume steady Pauli blocking (\ref{Paulirate}). However, these values are smaller than the maximum rate (\ref{maxrate}) computed in absence of Pauli blocking.

\section{Discussion}
\label{diss}

Our kinetic simulation reveals two physical effects in hot electrosphere, which were ignored in previous analyses. The first effect is the inflation of electrosphere due thermal evaporation of electrons, leading to its spatial extension to distances much larger than the electrostatic solution implies. The second effect is enhancement of the rate of pair creation due to pair simultaneous acceleration by the electric field. Both effects are crucial for estimation of pair creation rate, especially at low temperatures with strongly degenerate electrons, where analytical formulas fail to reproduce numerical rates, see Fig. \ref{fig0}. 

Our numerical results show that, as expected, the Schwinger process operates for nonzero temperature $T_S$. Positrons are accelerated in the Coulomb barrier and move outward. The total outward flux is approximately neutral due ability of electrons to overcome the Coulomb barrier. The distribution of the electric field and electron density is quasi static, and pair creation does not back react on the electrosphere. As a result electrons do not occupy all empty states and the process operates continuously.

In our simulations the interior of the compact object serves as a thermal bath of electrons. In the absence of collisions electrons with negative momentum are absorbed in the interior and equal number of electrons with positive momentum are created. Such mirror boundary conditions are adopted to ensure the energy and particle conservation on the finite computational grid. 
We do not investigate in this paper the fate of electrons moving inward, because we consider collisionless dynamics. It is expected that Coulomb collisions thermalize these electrons. 

We also do not discuss the thermal evolution of of the compact object and its impact on pair luminosity. It is expected that on sufficiently long timescale the temperature should decrease sufficiently to halt the Schwinger process. It has been argued by \citet{2002PhRvL..89m1101P} that neutrino cooling leads to rapid decrease of the surface temperature below MeV values. However, recently it was shown by \citet{2021ApJ...922..214L} that quark stars at high temperatures are opaque to neutrinos. Since neutrinos are captured within neutrinosphere, the cooling is less efficient.

Clearly on larger distances from the electrosphere pair outflow becomes collisional. Interaction between electrons and positrons eventually produces photons. Kinetic simulations of this process has been reported by \citet{2004ApJ...609..363A,2005ApJ...632..567A} assuming their rate is given by Usov's formula (\ref{usovrate}).

\section{Conclusions}
\label{concl}

In this paper we revisit Usov's mechanism for pair creation in electrosphere of compact astrophysical objects, such as hypothetical quark stars or neutron stars. As the density of electrons rapidly decreases outside the surface, electrosphere is essentially collisionless, and pair dynamics is governed by the Vlasov-Maxwell equations. Numerically solving these equations for hot electrosphere we found two effects, previously ignored in the literature. First, due to thermal evaporation of electrons electrosphere is inflated to much larger distances, though still microscopic, than electrostatic solution implies. Second, even for strongly degenerate distribution of electrons in electrosphere Pauli blocking is efficiently reduced by simultaneous acceleration of pairs created by the Schwinger process. Both these effects dramatically enhance pair creation rate, leading to luminosities that can be as large as $10^{52}$ erg/s, see eq. (\ref{maxrate}), much larger than previously derived.

These results imply that hot electrosphere may be a stronger source of relativistic pair outflows than previously assumed. Electron-positron pairs are generated in electrosphere in collisionless regime. 

{\bf Acknowledgements.} This work is supported within the joint BRFFR-ICRANet-2023 funding programme under the grant No. F23ICR-001.

\bibliography{total}{}
\bibliographystyle{aasjournal}

\end{document}